\documentclass[letterpaper]{article} 
\usepackage{url}  
\usepackage{graphicx}  
\usepackage{hyperref}
\usepackage{amsmath}
\usepackage{flushend} 
\usepackage{natbib}
\usepackage{apalike}
\usepackage{authblk}

\usepackage{color}
\usepackage[table]{xcolor}

\begin{document}

\title{Developments in the arms race between social bots and AI countermeasures} 
\title{Developments in AI to tell social bot and human behaviors apart} 
\title{On social bots, humans, and AI countermeasures}
\title{Arming the public with artificial intelligence to counter social bots} 

\author[1]{Kai-Cheng Yang}
\author[2]{Onur Varol}
\author[1]{Clayton A. Davis}
\author[3]{Emilio Ferrara}
\author[1,4]{Alessandro Flammini}
\author[1,4]{Filippo Menczer}
\affil[1]{Center for Complex Networks \& Systems Research, Indiana University}
\affil[2]{Center for Complex Networks Research, Northeastern University}
\affil[3]{Information Sciences Institute, University of Southern California}
\affil[4]{Indiana University Network Science Institute}

\date{}

\maketitle

\begin{abstract} 
The increased relevance of social media in our daily life has been accompanied by efforts to manipulate online conversations and opinions. Deceptive social bots --- automated or semi-automated accounts designed to impersonate humans --- have been successfully exploited for these kinds of abuse. Researchers have responded by developing AI tools to arm the public in the fight against social bots. Here we review the literature on different types of bots, their impact, and detection methods. We use the case study of Botometer, a popular bot detection tool developed at Indiana University, to illustrate how people interact with AI countermeasures. A user experience survey suggests that bot detection has become an integral part of the social media experience for many users. However, barriers in interpreting the output of AI tools can lead to fundamental misunderstandings. The arms race between machine learning methods to develop sophisticated bots and effective countermeasures makes it necessary to update the training data and features of detection tools. We again use the Botometer case to illustrate both algorithmic and interpretability improvements of bot scores, designed to meet user expectations. We conclude by discussing how future AI developments may affect the fight between malicious bots and the public.
\end{abstract}

\newpage

\section{Introduction}

During the past two decades, we have progressively turned to the Internet and social media to find news, share opinions, and entertain conversations  \citep{morris1996internet,smith2012twitter}. What we create and consume on the Internet impacts all aspects of our daily lives, including our political, health, financial, and entertainment decisions. This increased influence of social media has been accompanied by an increase in attempts to alter the organic nature of our online discussions and exchanges of ideas. In particular, over the past 10 years we have witnessed an explosion of social bots  \citep{lee2011seven,boshmaf2013design}, a presence that doesn't show signs of decline. 

Social bots are social media accounts controlled completely or in part by computer algorithms. They can generate content automatically and interact with human users, often posing as, or imitating, humans  \citep{ferrara2016rise}. Automated accounts can be harmless and even helpful in scenarios where they save manual labor without polluting human conversations. In this paper, however, we focus on those actors in online social networks that surreptitiously aim to manipulate public discourse and influence human opinions and behavior in an opaque fashion. 
While more traditional nefarious entities, like malware, attack vulnerabilities of hardware and software, social bots exploit human vulnerabilities, such as our tendencies to pay attention to what appears to be popular and to trust social contacts  \citep{jun2017perceived}. Unlike other social engineering attacks, such as spear phishing  \citep{jagatic2007social}, bots can achieve scalability through automation. For example, multiple accounts controlled by a single entity can quickly generate posts and make specific content trend or amplify misinformation. They can trick humans and engagement-based ranking algorithms alike, creating the appearance that some person or opinion is popular. Therefore, defending from social bots raises serious research challenges  \citep{boshmaf2012key}.


Manipulation of public opinion is not new; it has been a common practice since the dawn of humanity. The technological tools of all eras --- printed media, radio, television, and the Internet --- have been abused to disseminate misinformation and propaganda. The deceptive strategies employed on all these types of channels share striking similarities~\citep{varol2018deception}. Nevertheless, social media are particularly vulnerable because they facilitate automatic interactions via software. As a result, social media platforms have to combat a deluge of attacks. Facebook recently announced that 1.5 billion fake accounts were removed over six months in 2018.\footnote{\href{https://www.facebook.com/notes/mark-zuckerberg/a-blueprint-for-content-governance-and-enforcement/10156443129621634/}{A Blueprint for Content Governance and Enforcement. www.facebook.com/notes/mark-zuckerberg/a-blueprint-for-content-governance-and-enforcement/10156443129621634/}} Even a very low miss rate could leave millions of accounts available to be used as bots. In this light, it is not surprising that as many as 9--15\% of active Twitter accounts were estimated to be bots in 2017  \citep{varol2017online}, and that social bots are responsible for generating two thirds of links to popular websites \citep{pew2018bots}. 
Public interest in social bots has also dramatically increased during the past few years. 
A recent Pew survey shows that 66\% of Americans are aware of the existence of social bots and a consistent fraction of those believes that they are malicious \citep{pew2018social}. 
Despite high awareness, the survey also reveals that many people are not confident in their ability to identify social bots.  Part of the difficulty in learning to recognize bots is that they continuously evolve in response to efforts by platforms to eliminate them. 
Social bots that go undected by platforms have become increasingly sophisticated and human-like  \citep{varol2017online}, introducing an arms race between bot creators and the research community. Indeed, industry and academic research about social bots has flourished recently.

A first concern of this paper is to briefly review  recent studies about social bots and the development of AI countermeasures.
A second goal is to present a study on how people interact with bot detection tools and what insight they gain. This is an important perspective that is often neglected by developers of AI tools. After all, winning this fight requires broad participation by social media users.

\section{Literature review}

There are many types of social bots. In this section, we review recent studies on the diverse characteristics of social bots, their activities, and impact. We then review the literature on bot detection algorithms. 

\subsection{Characterization of social bots}

Some simple bots only do one thing: they post content automatically.
The Twitter account \texttt{@big\_ben\_clock} is a quintessential representative of this class --- it tweets every hour mimicking the real Big Ben. 
Similar accounts can automatically post or retweet/share to publicize information such as news and academic papers  \citep{lokot2016news,haustein2016tweets}. 
These bots are naive and easy to identify: they share only one type of content and they don't try to misrepresent themselves or their motivations. 
In 2011, \cite{lee2011seven} identified thousands of social bots. 
They created bots that posted meaningless messages through automated scripts. The underlying expectation was that only other bots and no human  would follow these honeypot accounts. The honeypot accounts did end up having many followers. Subsequent analysis confirmed that these followers were indeed bots and revealed the common strategy of randomly following accounts to grow the number of social connections. 

More sophisticated bots adopt various strategies to impersonate human users. 
Social bot developers can populate bot profiles by searching and collecting material from other platforms.
A more extreme example of these kinds of bots are identity thieves: they copy usernames, profile information, and pictures of other accounts and use them as their own, making only small changes  \citep{cresci2015fame}. 
State-of-the-art machine learning technologies can be employed as part of the algorithms that automatically generate bot content  \citep{freitas2015reverse}. 
Sophisticated bots can emulate temporal patterns of content posting and consumption by humans. 
They can even interact with other users by engaging in conversations, commenting on posts, and answering questions  \citep{hwang2012socialbots}.

Bots employ a variety of strategies to form an audience. 
There are bots designed to gather followers and expand social circles, with the goal of exerting some form of influence.
Some bots, for example, search the social networks for popular accounts, follow them, and ask to be followed back  \citep{aiello2012people}.
This type of infiltration has been proven to be more effective when the target community is topically centered. 
Bots can identify and generate appropriate content around specific topics to gain trust and attention from people interested in those topics. Such attention often translates in following relationships  \citep{freitas2015reverse}. 
This finding is even more worrisome when one considers that the targets were programmers, a population expected to be more aware of social bots compared to other users. 

Another class of bots, whose activity is less evident, is that of so-called fake followers  \citep{cresci2015fame,de2013twitter}.
These accounts are managed by entities that get paid in exchange for following customers who want to increase their perceived popularity. 
The fake followers often follow each other, forming a network that lends credibility to each member and allows them to elude being flagged for lack of followers. 
A few dollars can buy thousands of fake followers. 
A recent report\footnote{\href{www.nytimes.com/interactive/2018/01/27/technology/social-media-bots.html}{The Follower Factory. www.nytimes.com/interactive/2018/01/27/technology/social-media-bots.html}} revealed that reality television stars, professional athletes, comedians, TED speakers, pastors, and fashion models all purchased fake followers on Twitter at least once.
The companies that sell fake followers also sell fake engagement (likes, comments, retweets) and even custom-made fake personas.

A group of social bots, sometimes referred to as a botnet, may act in coordination, posting the same content again and again to generate false popularity. 
This type of behavior is generally harder to detect as the accounts involved, when inspected individually, may appear as genuine. Their behavior becomes conspicuous only when one is able to detect a large number of such accounts acting in a strongly coordinated manner  \citep{chavoshi2016debot}. Not surprisingly, hybrid approaches that mix human and automated activity have been devised to bypass bot and bot-coordination detection systems 
 \citep{grimme2018changing}. 
A recent study by \cite{cresci2017paradigm} shows that neither humans nor supervised machine learning algorithms can identify this kind of bots successfully. 

\subsection{Activity and impact of social bots}
\label{sec:impact}

Bot activity has been reported in several domains, with the potential to affect behaviors, opinions, and choices. 
Health is one domain of particular concern \citep{allem2018could}, where we have observed social bots influencing debates about vaccination policies \citep{ferrara2016rise,broniatowski2018weaponized} and smoking \citep{allem2016importance,allem2017cigarette}. Politics is another key domain. During the 2010 U.S. midterm elections, primitive social bots were found to support some candidates and attack their opponents \citep{metaxas2012social}, injecting thousands of tweets pointing to websites with fake news \citep{ratkiewicz2011detecting}.

During the 2016 U.S. presidential election, social bots were found to generate a large amount of content, possibly distorting online conversations. Our analysis \citep{bessi2016social} suggests that retweets were a vulnerable mode of information spreading: there was no significant difference in the amount of retweets that humans generated by resharing content produced by other humans or by bots. In fact, humans and bots retweeted each other substantially at the same rate. This suggests that bots were very effective at getting messages reshared in the human communication channels. 
We further explored how bots and humans talked about the two presidential candidates. We noted that bots tweeting about Donald Trump generated the most positive tweets \citep{bessi2016social}. The fact that bots produce systematically more positive content in support of a candidate can bias the perception of the individuals exposed to this content, suggesting that there exists an organic, grassroots support for a given candidate, while in reality it is all artificially generated. 

Similar cases of political manipulation were reported in other countries \citep{stella2018bots}. Our analysis highlighted the presence of Twitter bots during the 2017 French Presidential Election \citep{ferrara2017disinformation}. 
We explored tweets related to candidates Marine Le Pen and Emmanuel Macron in the days leading to the election. We identified 18K active bots that pushed the so-called \textit{MacronLeaks} disinformation campaign. Accounts who engaged with these bots were mostly US-based users with pre-existing interest in alt-right topics and alternative news media, rather than French users. Hundreds of bots posting about the French elections were also active in the 2016 US Election discussion, suggesting the existence of a black market for reusable political disinformation bots.

The spread of fake news online is another area in which the effect of bots is believed to be relevant  \citep{lazer2018science,bessi2015science}. 
A study based on 14 million tweets posted during and after the 2016 U.S. presidential election revealed that bots played a key role in the spread of low-credibility content \citep{shao2018spread,shao2018anatomy}. The study uncovered strategies by which social bots target influential accounts and amplify misinformation in the early stages of spreading, before it becomes viral. 
Even bots created with good intentions may contribute to the spread of misinformation. After the Boston marathon bombing, for example, bots started to automatically retweet indiscriminately, without verifying  the credibility of the posts being retweeted or that of their sources \citep{gupta20131}, unintentionally amplifying the diffusion of misinformation. 

Social bots have been used to promote terrorist propaganda and proselytize online extremism. 
By analyzing a sample of 20,000 ISIS-supporting accounts, \cite{berger2015isis} found that that the terrorist organization was actively using social media and bots to spread its ideology.
\cite{abokhodair2015dissecting} dissected a botnet that tried to misdirect the online discussion during the Syrian civil war in 2012. 
\cite{ferrara2016predicting} analyzed a dataset of millions of extremist tweets and found that social bots produced some of the content. 

\subsection{Bot detection methods}
\label{sec:review-methods}

An early attempt by \cite{wang2012social} involved building a crowdsourcing social bot detection platform. 
This method proved to be effective, but scalability was a prominent issue: while bots can be multiplied at will at essentially no cost, human detectors cannot. Later efforts therefore mostly leveraged machine learning methods.

Approaches based on supervised machine learning algorithms are the most common \citep{subrahmanian2016darpa}.
Supervised approaches depend and often start with the collection of an extensive dataset, with each account labeled as either human or bot.
These labels usually come from human annotation  \citep{varol2017online}, automated methods (\textit{e.g.}, those based on honey pots  described earlier \citep{lee2011seven}), or botnets that display suspicious behaviors  \citep{echeverria2017discovery2,echeverria2017discovery1}. 
A critical issue with existing datasets is the lack of ground truth.  There is no objective, agreed-upon, operational definition of social bot. One of the factors that explain this is the prevalence of accounts that lie in the gray area between human and bot behavior, where even experienced researchers cannot easily discriminate.
Nevertheless, datasets do include many typical bots; using the training labels as proxies for ground truth  
makes it possible to build practically viable tools \citep{davis2016botornot,varol2017online}. 

The choice of relevant features used to describe entities to be classified is a critical step of machine learning classifiers. Different choices have been considered, but in general six broad categories of features have been identified as relevant for discriminating between human and bot accounts \citep{varol2018feature}: user metadata, friend metadata, retweet/mention network structure, content and language, sentiment, and temporal features. In the case of supervised learning, after extraction and preprocessing, the features are fed into supervised machine-learning models for training, then the trained models are used to evaluate previously unseen accounts.

Most techniques attempt to detect bots at the \textit{account} level, by processing many social media posts to extract the features listed above. Recently, \cite{kudugunta2018deep} proposed a deep neural network approach based on the contextual long short-term memory (LSTM) architecture that exploits content and metadata to detect bots at the \textit{tweet} level. Contextual features are extracted from user metadata and fed as auxiliary input to LSTM deep nets processing the tweet text. The proposed framework, designed to predict whether a given tweet was posted by a human or a bot, exhibited promising performance on test benchmarks. 

While supervised methods have proven to be effective in many cases, they do not perform well at detecting coordinated social bots that post human-generated content \citep{cresci2017paradigm,grimme2018changing,chen2018unsupervised}.
As mentioned earlier, those coordinated bots are not usually suspicious when considered individually.
Their detection requires information about their coordination, which becomes available only once the activity of multiple bots is considered.
Unsupervised learning methods have been proposed to address this issue. Early attempts combined multiple features via Euclidean distance and then applied clustering or network community algorithms  \citep{ahmed2013generic,miller2014twitter}.
\cite{cresci2016dna} encoded tweet type or content as strings of symbols, then considered the longest common substring to detect coordination.
By comparing the time series of accounts sampled from the Twitter streaming API, \cite{chavoshi2016debot} built an unsupervised tool called DeBot that is able to find accounts tweeting in synchrony, suggesting they are automated.
\cite{chen2018unsupervised} adopted a similar method to detect bots by finding accounts tweeting similar content. 
These unsupervised methods have the advantage of focusing on what accounts (or tweets) have in common rather than what is distinct, but also the disadvantage of having to consider quadratic numbers of account pairs in the worst case.

A recent research direction is to test the limits of current bot detection frameworks in an adversarial setting. The idea is to propose methodologies to engineer systems that can go undetected.  \cite{cresci2019capability} proposed the use of evolutionary algorithms to improve social bot skills. 
\cite{grimme2018changing} employed a hybrid approach involving automatic and manual actions to achieve bots that would be classified as human by a supervised bot detection system. 
Despite the good intention of pointing to weaknesses in existing systems, this research might also inspire bot creators and give them a competitive advantage.

\section{User engagement with bot detection tools}
\label{sec:engagement}

Research efforts aimed at bot detection may be valuable to mitigate the undesired consequences of bot activity, but their efficacy is limited by two factors: limited public awareness of the bot problem, and unwillingness to adopt sophisticated tools to combat it.
The success of research efforts also critically depends on the capacity to adapt to ever changing and increasingly sophisticated artificial accounts. This, in turn, depends on the ability to engage the public and collect the feedback it provides. It is therefore important to understand how the public adopts, interacts with, and interprets the results of bot detection tools. This perspective is often neglected by research on social bots.
In this section, we use Botometer --- a bot detection tool developed at Indiana University --- as a case study to illustrate efforts aimed at understanding the needs of the users and how their feedback can help improve tools. 

Botometer is based on a supervised machine learning approach \citep{davis2016botornot,varol2017online}. Given a Twitter account, Botometer extracts over 1,000 features relative to the account from data easily provided by the Twitter API, and produces a classification score called \textit{bot score}: the higher the score, the greater the likelihood that the account is controlled completely or in part by software, according to the algorithm. 
Because some of the features are based on English-language content, the bot score is intended to be used with English-language accounts. To evaluate non-English accounts, Botometer also provides a language-independent score that is produced by a classification model trained excluding linguistic features. 
Botometer additionally reports six subscores, each produced by a model based on a distinct subset of features. The score names refer to the feature classes: user meta-data, friends, content, sentiment, network, and timing. 
The subscores are provided to help users identify which features contribute to the overall score. 

\begin{figure*}
    \centering
    \includegraphics[width=\textwidth]{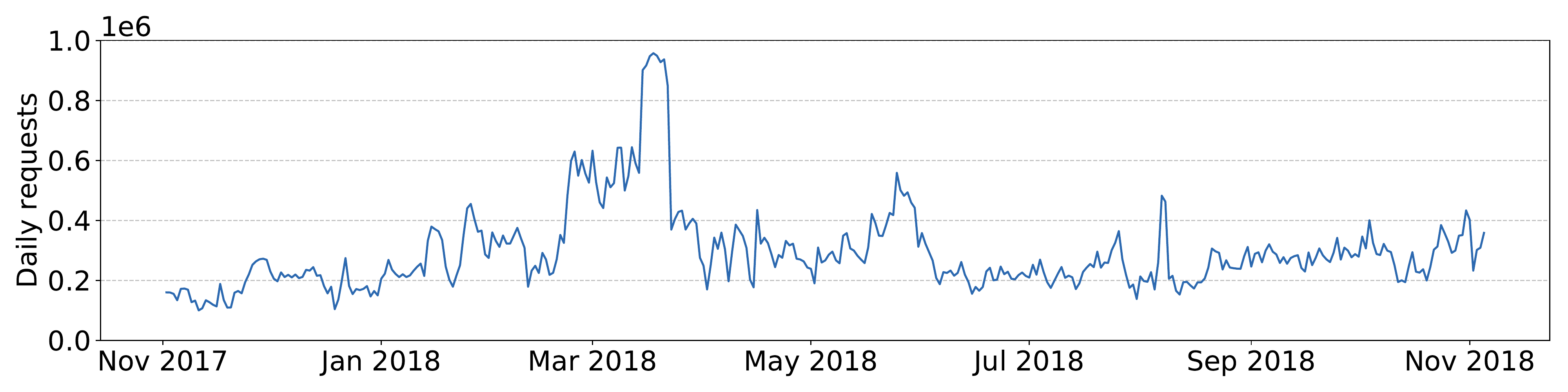}
    \caption{Number of daily requests handled by Botometer.}
    \label{fig:botometer_flux}
\end{figure*}

The earliest version of Botometer (known as ``BotOrNot'') became available to the public in May 2014. Free access is offered through both a web interface (\url{botometer.org}) and an API. In the past four years, Botometer has constantly increased its basin of adoption, and has provided data for a number of influential studies, including by \cite{vosoughi2018spread} and the Pew Research Center \citep{pew2018bots}. Even more importantly, Botometer is used by many regular Twitter users. Currently, it handles over a quarter million requests every day (Figure~\ref{fig:botometer_flux}), while the website receives over 500 daily visits. Botometer also supports several third-party tools, including browser extensions, bot-hunting bots, and localized versions in non-English-speaking countries.

\begin{figure}
    \centering
    \includegraphics[width=\textwidth]{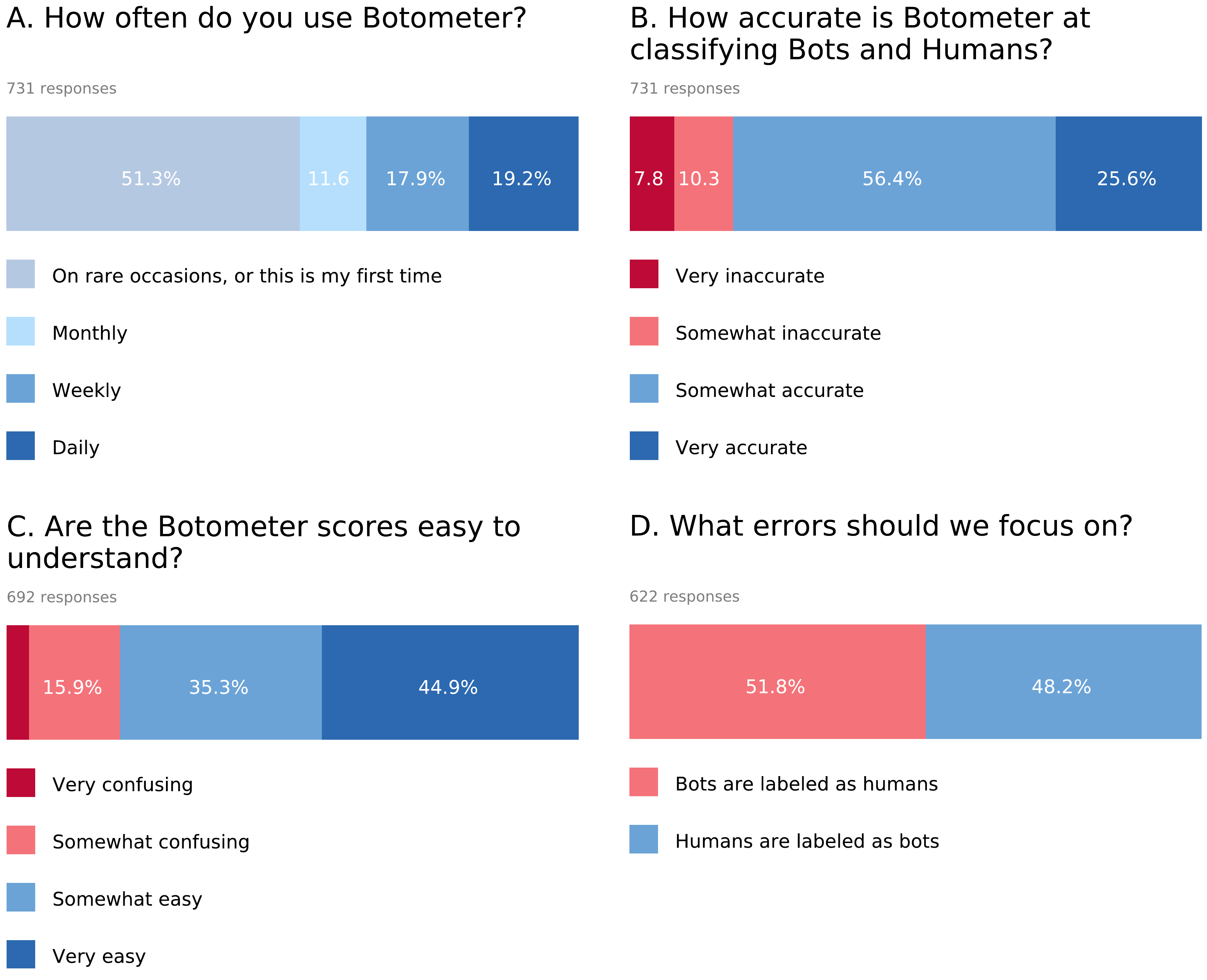}
    \caption{Botometer user experience survey questions and responses. Questions A and B are required, others are optional.}
    \label{fig:botometer_ue_survey}
\end{figure}

\begin{figure}
\centering
\includegraphics[width=0.7\columnwidth]{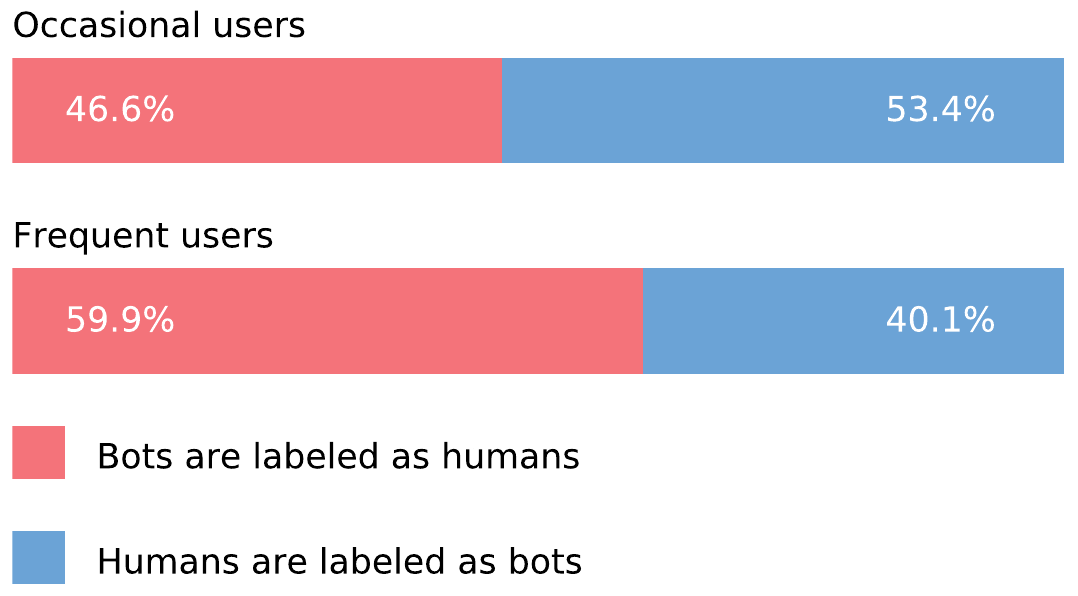}
\caption{Responses to the Botometer user experience survey question about error type concerns (Figure~\ref{fig:botometer_ue_survey}D), grouped by frequency of usage. Respondents who use Botometer daily or weekly are considered frequent users.}
\label{fig:botometer_errortype_freq}
\end{figure}

Between August and October 2018, we conducted a user experience survey with participants recruited among the visitors of the Botometer website. The survey contained two questions with required answers and a few optional ones. It also allowed respondents to enter some free-text comments. We collected usable answers from 731 participants; they are listed in Figure~\ref{fig:botometer_ue_survey} together with the questions. A few interesting facts emerge. First, more than one third of participants 
use Botometer at least weekly, implying that detecting bots is becoming a recurring need for at least some users (Figure~\ref{fig:botometer_ue_survey}A).
Second, over 80\% of the users believe Botometer is accurate in classifying bots and humans (Figure~\ref{fig:botometer_ue_survey}B). 
Third, over 80\% of the users find the bot scores presented by Botometer easy to understand  (Figure~\ref{fig:botometer_ue_survey}C).
Although these numbers are encouraging, we are aware of self-selection bias, as respondents of the survey tend to be active users of Botometer. 
Finally, users seem to be equally worried by false positives (humans misclassified as bots) and false negatives (bots misclassified as humans) as shown in Figure~\ref{fig:botometer_ue_survey}D. 
A more nuanced picture of this last observation emerges when these results are disaggregated into frequency-of-usage categories: Figure~\ref{fig:botometer_errortype_freq} shows that occasional users care more about false positives. A typical example of usage that might elicit this concern is when an individual checks whether their own account looks like a bot. On the other hand, frequent users care more about false negatives. For example, an advertiser interested in the number of fake followers of a celebrity paid to endorse a product may be more worried about missing bots.  

\begin{figure}
\centering
\includegraphics[width=0.7\columnwidth]{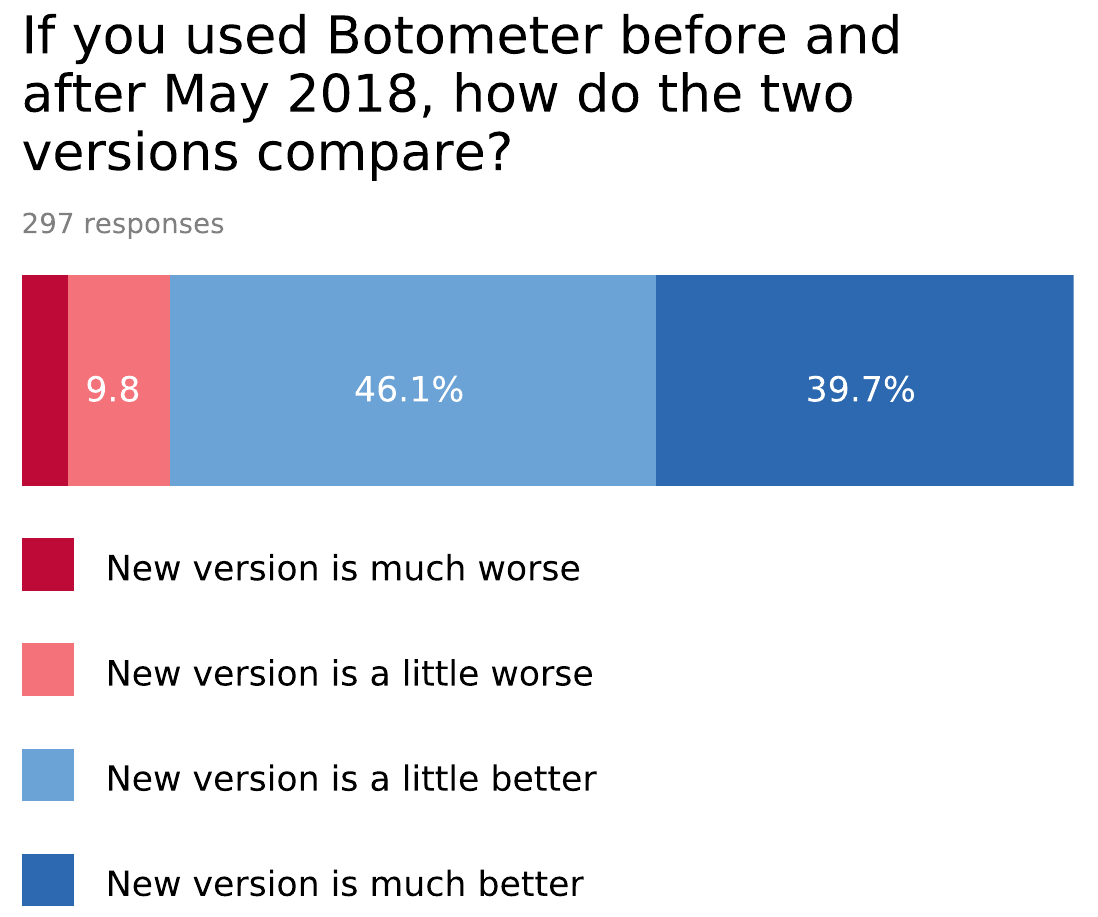}
\caption{Botometer user experience survey result: comparison between Botometer v2 and v3.}
\label{fig:botometer_version}
\end{figure}

Botometer is based on Random Forest classifiers \citep{breiman2001random}. The score produced by such a model is the fraction of trees that classify the account under examination as a bot. 
This interpretation of the bot score is obscure to the typical user, who is unlikely to have background knowledge in machine learning. Because a bot score is defined in the unit interval, it is tempting to interpret it as the \textit{probability} that the account is a bot --- if the score is 0.3, say, then 30\% of accounts with similar feature values are bots. Another interpretation is that such an account is 30\% automated. 
Judging from early user feedback, such incorrect interpretations were prevalent. Despite an explanation offered on the website, the confusion persisted. 
To address this concern and bridge the gap between classifier output and user expectations, in the next section we describe  three changes in how scores are presented: a non-linear re-calibration of the model, a linear rescaling of scores to the $[0,5]$ range, and the introduction of a new statistics called Complete Automation Probability. These changes were incorporated into the latest version of Botometer (v3), launched in May 2018, together with other improvements described in Section~\ref{sec:retraining}. 
Users seem to appreciate these changes, as shown in Figure~\ref{fig:botometer_version}. 

\section{Bot score interpretability}

\subsection{Model calibration}

As previously described, the raw output of the random forest classifier is not a probability. Consider the classifier $C$ and two Twitter accounts $A$ and $B$ such that $C(A) = x$ and $C(B) = y$ with $x < y$. Here we can say that account $A$ is less likely to be a bot than account $B$, but we cannot say that account $A$ is $x\%$ likely to be a bot. While this is not a problem for measures like AUC, which quantifies the probability that a classifier will rank a randomly chosen bot higher than a randomly chosen human, it does present an obstacle when interpreting bot scores.

Proper contextualization of the raw classifier output requires comparison to the rest of the scores in the training set. For example, maximum-likelihood estimation is employed by \cite{varol2017online} to find the appropriate cutoff for binary classification: the best score above which an account is classified as a bot and below which it is classified as human. This threshold approach is useful for scientists counting bots in a collection  \citep{suarez2016influence,haustein2016tweets,gorwa2017computational,stella2018bots,shao2018spread}, but it is not intuitive for humans evaluating a single account.

A \textit{calibrated} classifier $C'$ is one whose output can be  interpreted as a probability, \textit{i.e.}, one in which $C'(A) = x$  means that the classifier estimates as $x\%$ the probability that account $A$ is a bot. As a corollary, the binary classification threshold of a well-calibrated model is 50\%. Calibrated models are ideal from the point of view of interpreting their results. To generate such a calibrated model $C'$ from our classifier $C$, we want a calibration function $F$ such that $F \circ C = C'$. 

\begin{figure}
    \centering
    \includegraphics[width=0.48\linewidth]{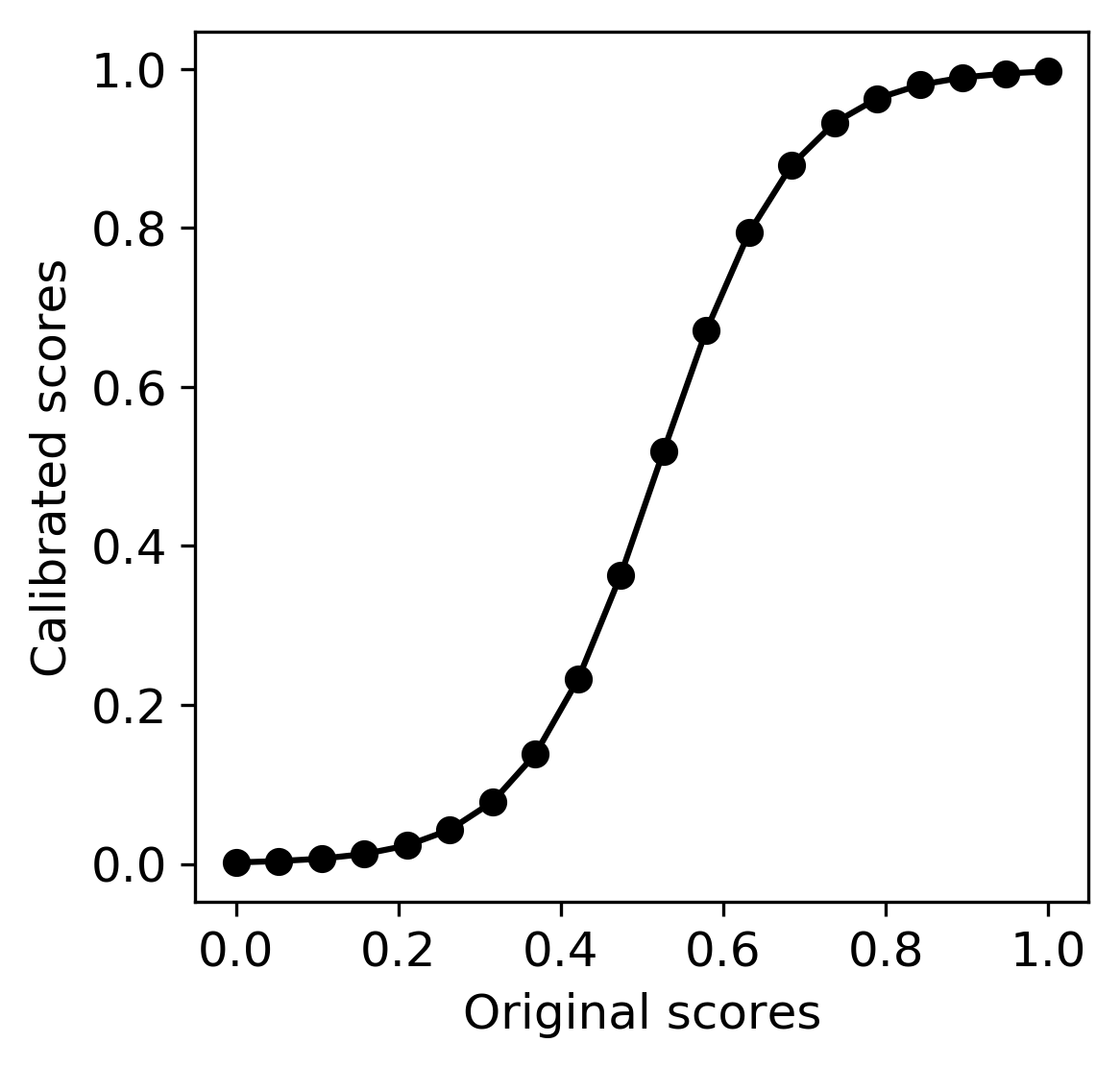}
    \includegraphics[width=0.48\linewidth]{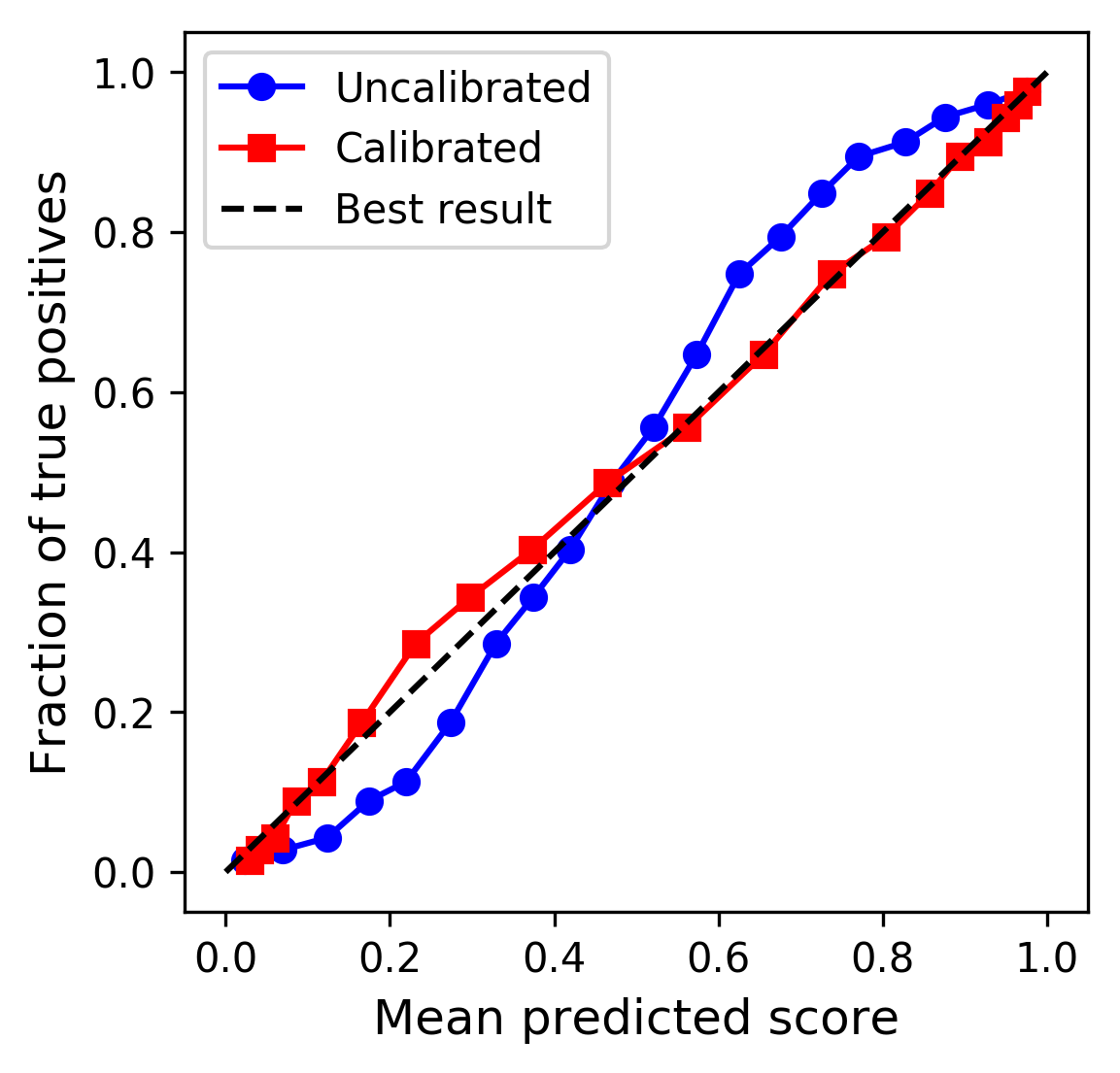}
    \caption{Calibration of the bot scores. The mapping function projects raw classifier outputs to calibrated scores (left). Reliability curves plot true positive rates against mean predicted scores (right). The calibrated curve indicates higher reliability because it is closer to the unbiased diagonal line.}
    \label{fig:botscore_calibration}
\end{figure}

To calibrate Botometer, we employed Platt's scaling  \citep{niculescu2005predicting}, a logistic regression model trained on classifier outputs.
Figure~\ref{fig:botscore_calibration} presents the initial model outcomes and the calibrated scores. 
Note that this mapping shifts scores within the unit interval but preserves order, therefore leaving the AUC unchanged. The figure also shows reliability diagrams for raw and calibrated scores  \citep{degroot1983comparison}. 
We split the unit interval into 20 bins; an instance in the training dataset is assigned to a bin based on its predicted (raw) score. 
For each bin, the mean predicted score is computed and compared against the fraction of true positive cases.
In a well-calibrated model, the points should align with the diagonal.
Observe that the blue line on the right side of Figure~\ref{fig:botscore_calibration} is steepest in the middle of the range; this is because most uncalibrated bot scores fall near the middle of the unit interval. Since users, when presented with a single uncalibrated bot score, do not know that most scores fall into this relatively narrow range, they are misled into perceiving uncertainty about classification of most accounts. The flatter red line in the plot shows that each bin has approximately the same number of scores in the calibrated model.

As mentioned earlier, the percentage of bot accounts is estimated around 9--15\%  \citep{varol2017online}, which means the probability of a randomly selected account being automated is low. 
The calibrated score cannot reflect this fact, since it is a likelihood estimation based on our roughly balanced (half humans, half bots) training datasets.
In the next subsection we offer a Bayesian posterior probability that overcomes this problem. 
The bot score presented on the current Botometer website (v3) is the calibrated likelihood score, linearly transformed to a $[0, 5]$ scale in order to differentiate it from the probability that an account with that score is a bot. 
The API, whose users tend to be more advanced, also provides uncalibrated scores. 

\subsection{Complete automation probability}

Users of Botometer are asking the question: ``Is this account a bot, or not?'' While model calibration makes the bot scores easier to interpret, a calibrated model still lacks one additional piece of information necessary to answer this ``bot or not'' question: the background level of bots in the population. 
To see why, consider that the less likely a claim, the more evidence is required to accept it. In our case, the rarer the bots, the higher the score needed to have confidence that an account is a bot.  
%
%
This connection between likelihood and background probability is formalized by Bayes' theorem. 
The  ``bot or not'' question can be expressed mathematically as $P(\textrm{Bot} \mid S)$, the conditional probability that an account is a bot given its bot score $S$. Applying Bayes' rule allows us to rewrite the conditional probability as:
\begin{equation}
   P(\textrm{Bot} \mid S) = P(\textrm{Bot})
  \frac{P(S \mid \textrm{Bot})}{P(S)}.
  \label{eq:bayes}
\end{equation}
We call $P(\textrm{Bot} \mid S)$, the quantity we are after, the \textit{posterior} probability. The \textit{prior} probability $P(\textrm{Bot})$ is the background probability that any randomly-chosen account is a bot.
The $\frac{P(S \mid \textrm{\scriptsize Bot})}{P(S)}$ term is called the \textit{evidence}. It compares the \textit{likelihood} that a bot has score $S$, $P(S \mid \textrm{Bot})$, with the probability that any account has that score. Since this is a ratio of functions of the score $S$, it is not affected by transformations of $S$ as done in model calibration. Therefore we use raw, uncalibrated scores in our calculations. 

\begin{figure}
    \centering
    \includegraphics[width=\textwidth]{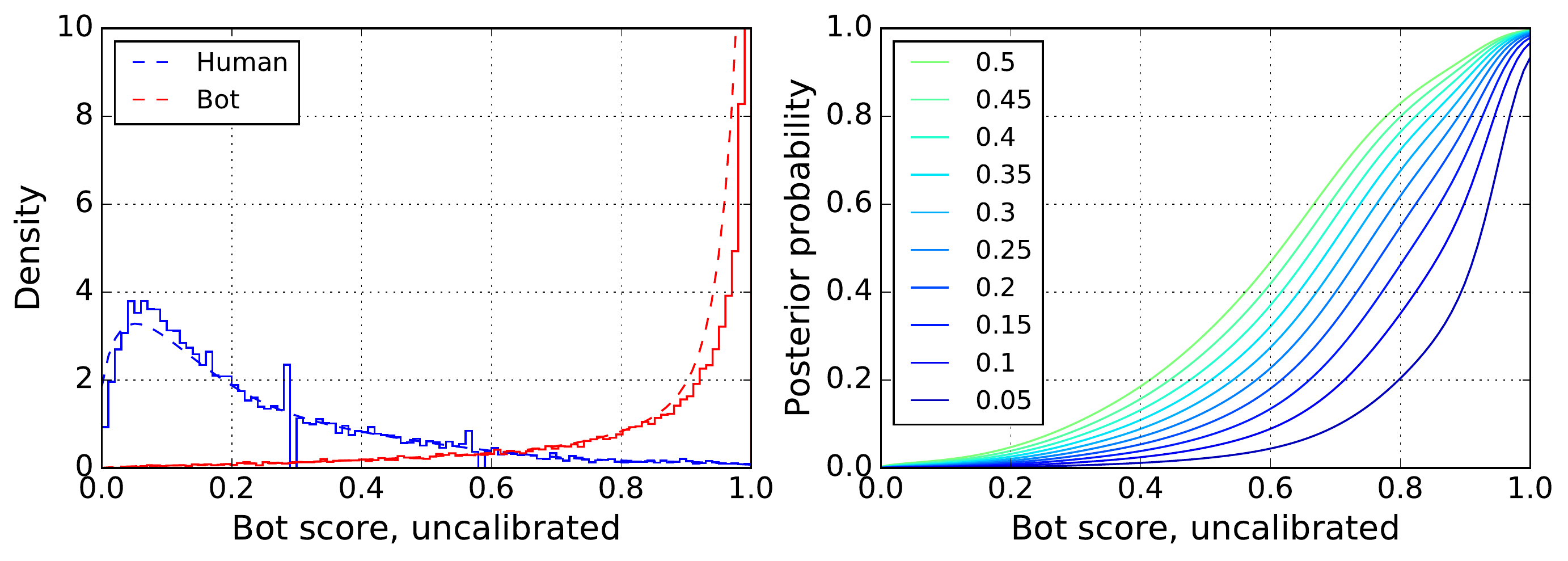}
    \caption{Likelihood distributions (left) and posterior probabilities (right) for the CAP calculation. The left plot shows the binned, empirical bot score distribution for accounts labeled human and bot, along with dashed lines displaying the density estimate for each. On the right, posterior probability curves are calculated for several choices of the prior, $P(\textrm{Bot})$.}
    \label{fig:CAP}
\end{figure}

To calculate the posterior probability, we can expand the denominator in the evidence term of Eq.~\ref{eq:bayes}:
\begin{equation}
    \begin{split}
    P(S) & = P(S \mid \textrm{Bot})P(\textrm{Bot})
    + P(S \mid \textrm{Human})P(\textrm{Human}) \\
         & = P(S \mid \textrm{Bot})P(\textrm{Bot})
    + P(S \mid \textrm{Human})(1 - P(\textrm{Bot})).
    \end{split}
\end{equation}
The task is then to obtain distributions for the likelihoods $P(S \mid \textrm{Bot})$ and $P(S \mid \textrm{Human})$. The model's training data provides empirical distributions of scores for both humans and bots (Figure~\ref{fig:CAP}); density estimation can be used to find a probability density function likely to produce a particular empirical distribution. Binning is the simplest approach to density estimation, sometimes employing a sliding window. However, this approach proved unsuitable because quantization artifacts in the Botometer classifier output lead to discontinuities in the density functions.

The existence of these artifacts suggests curve-fitting or kernel density estimation (KDE) methods. KDE is widely used for generating distribution functions from empirical data, but the commonly-used implementations assume unbounded support, \textit{i.e.}, a theoretically infinite range of possible bot score values. Since the support of our distribution is necessarily the unit interval, the use of KDE in this case leads to biased estimates near the interval boundaries \citep{leblanc2012}.

Seeking to avoid this systematic bias, Botometer uses curve fitting, choosing among curves that naturally admit a bounded support. One well-known candidate meeting this criterion is the Beta distribution. However, three different fitting algorithms (maximum-likelihood, method of moments, and negative log-likelihood) all failed to produce adequate fits. This is likely due to the shape of the $P(S \mid \textrm{Human})$ curve, with its mode near, but not at, zero. 
Ultimately we chose a lesser-known family of curves for our fit: Bernstein polynomials \citep{babu2002}. 
Implementation is straightforward as computing values of a polynomial curve is simple, even one with high degree. Experiments showed that a 40th-degree polynomial fit produced satisfactory results (Figure~\ref{fig:CAP}).

With likelihood curves generated from our classifier testing data, it is straightforward to calculate the evidence term in Bayes' rule. The other term is the prior $P(\textrm{Bot})$, the background probability of a given account being a bot. This is necessary because the training data does not include information on how common bots are in the wild. Failure to take the prior into account leads one to overestimate the number of bots in a sample due to humans being more common than bots, leading to false positives outnumbering false negatives. The best we can do is use an informed estimate for the prior.
In general, one could allow users to estimate the prior probability; a user might have information about the proportion of bots in a particular sample. For example, a sample of suspicious accounts would require a high prior probability.
Botometer uses $P(\textrm{Bot})=0.15$, corresponding to a previously published estimate of the proportion of bots on Twitter  \citep{varol2017online}. 

Taken together, the prior and the evidence allow us to calculate the posterior probability that an account with a particular bot score $S$ is a bot. In Botometer, the posterior is called Complete Automation Probability (CAP). As shown in Figure~\ref{fig:CAP}, this probability is generally more conservative than the calibrated bot score, reflecting the relative rarity of bots.
The reported CAP estimates the probability that the account is indeed a bot, and gives end users the information they need to act on the data returned. For example, there are Twitter bots that use Botometer to call out other bot accounts. Whereas these bot-hunting bots previously had to pick a somewhat-arbitrary threshold, now they can use the CAP to make an informed decision based on their tolerance for the risk of falsely labeling accounts.

\section{Updating detection models}
\label{sec:retraining}

A supervised machine learning tool is only as good as the data used for its training. 
Social bots evolve rapidly, and even the most advanced algorithms will fail with outdated training datasets.
Therefore it is necessary to update classification models, using newly available data as well as feedback collected from  users.
At the same time, one must continuously evolve the set of features that may discriminate between human behaviors and increasingly complex bot behaviors. 
Let us again use the Botometer case to illustrate the need for retraining bot detection models via new data and feature engineering. 


\begin{table*}
\small
\begin{tabular}{lrrp{5cm}}
    \hline
    Dataset name & \#bots & \#human & Notes\\
    \hline
    \texttt{caverlee} & 22,179 & 19,276 & Honeypot-lured bots and sample human accounts  \citep{lee2011seven}\\
    \texttt{varol-icwsm} & 826 & 1,747 & Manually labeled bots and humans sampled by Botometer score deciles  \citep{varol2017online}\\
    \texttt{cresci-17} & 10,894 & 3,474 & Spam bots and normal humans  \citep{cresci2017paradigm}\\
    \texttt{pornbots} & 21,963 & 0 & Pornbots shared by Andy Patel (\url{github.com/r0zetta/pronbot2})\\
    \texttt{celebrity} & 0 & 5,970 & Celebrity accounts collected by CNetS team \\
    \texttt{vendor-purchased} & 1,088 & 0 & Fake followers purchased by CNetS team \\
    \texttt{botometer-feedback} & 143 & 386 & Botometer feedback accounts manually labeled by author K.-C.Y. \\
    \texttt{political-bots} & 62 & 0 & Automated political accounts run by \texttt{@rzazula} (now suspended), shared by \texttt{@josh\_emerson} on Twitter \\
    \hline
    Total & 57,155 & 30,853 & All datasets available at \url{botometer.iuni.iu.edu/bot-repository} \\
    \hline
\end{tabular}
    \caption{Training data used by Botometer. CNetS refers to the Center for Complex Networks and Systems Research at Indiana University.}
    \label{table:trainingdata} 
\end{table*}

\begin{table*}
\small
\begin{tabular}{p{2.5cm}p{1.5cm}p{3cm}p{3.2cm}}
    \hline
    Version \\(activity period) & \#features & Training datasets & Notes\\
    \hline
    \textbf{v1} (2014-05-01 -- 2016-05-03) & 1150 & \texttt{caverlee} & Initial version of Botometer (known as BotOrNot at the time) \\
    \textbf{v2} (2016-05-03 -- 2018-05-11) & 1150 & \texttt{caverlee}, \texttt{varol-icwsm} & Version used in results presented by  \cite{varol2017online}\\ 
    \textbf{v3} (2018-05-11 -- present) & 1209 & \texttt{caverlee}, \texttt{varol-icwsm}, \texttt{cresci-17}, \texttt{pornbots}, \texttt{vendor-purchuased}, \texttt{botometer-feedback}, \texttt{celebrity} & New features introduced to capture more sophisticated behaviors and to comply with Twitter API changes.\\
    \hline
\end{tabular}
    \caption{Datasets and numbers of features used for training subsequent versions of Botometer models.}
    \label{table:botometer_version} 
\end{table*}

In the past few years, several bot datasets collected by colleagues in academia have been included in our training data. 
Such data are now shared with the public at large via a platform called \emph{Bot Repository} (\url{botometer.iuni.iu.edu/bot-repository}). 
Table~\ref{table:trainingdata} lists the datasets currently included and used for training, while Table~\ref{table:botometer_version} summarizes the different versions of Botometer and the growing training dataset it employed.


Retraining does not only mean adopting new datasets as they become available, but also engineering new features --- or redesigning features that are already in use --- to best describe the behaviors of new classes of bots.
In light of journalistic evidence and other academic research, the list of features employed by Botometer was recently enriched to incorporate new ones that are designed to capture bots employed in information operations. Examples include:
\begin{itemize}
    \item{\textbf{Time zones:}} an account would be suspicious if its profile indicates the US Eastern Standard Time zone while most of its followers appear to be in the Moscow Standard Time zone. 
    \item{\textbf{Language metadata:}} anomalous patterns in language use and audience language preferences can be revealed by a low fraction of neighbors (friends, followers) having the same language as a target account.
    \item{\textbf{Device metadata:}} we capture the types of devices and platforms used for posting tweets, as well as the entropy across different platforms. 
    \item{\textbf{Content deletion patterns:}} highly-active accounts frequently create and delete content to hijack user attention without revealing too much information on their excessive posting behavior. We incorporated features to estimate content production rates based on the account creation date, metadata on total tweets, recent content volume, and inter-event times between recent posts --- the mismatch between these statistics can be used as a proxy of content deletion. An account that appears to have tweeted once per day on average over the past five years, but has generated 500 tweets per day during the last two days, should raise flags. 
\end{itemize}

\begin{figure}
    \centering
    \includegraphics[width=\textwidth]{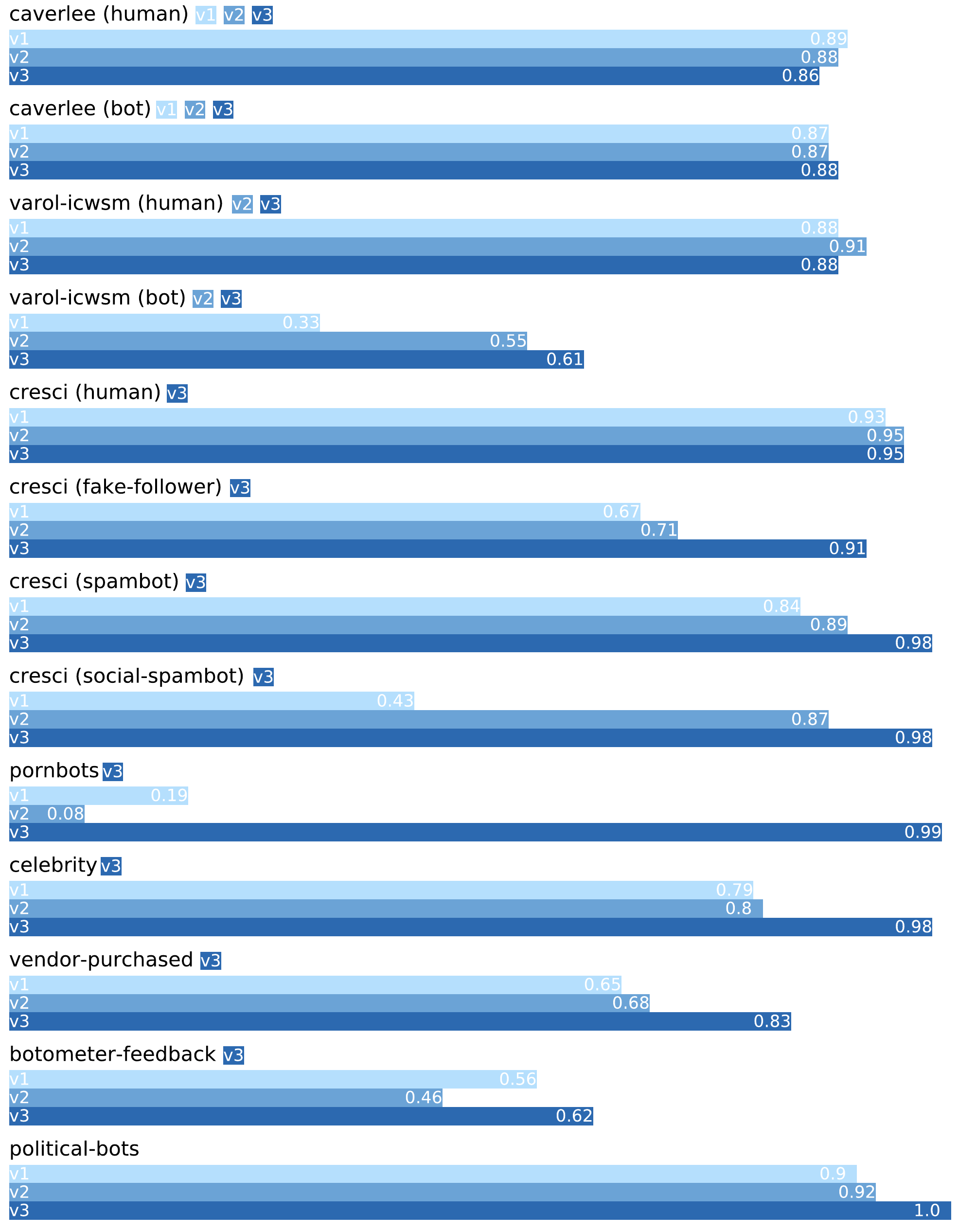}
    \caption{Model accuracy (percentage of instances correctly classified) based on in-sample evaluation via cross-validation and testing on out-sample datasets. In-sample evaluation is indicated by the version tags, which list the versions of Botometer in which each dataset was used for training. See Tables~\ref{table:trainingdata} and \ref{table:botometer_version} for further information on datasets and Botometer versions, respectively.}
    \label{fig:model_performances}
\end{figure}

When updating a supervised learning framework for bot detection, a central question is whether, or to what extent, the retrained model can recognize new classes of bots on which the model has not been trained. 
To explore this question about generalization, we trained models on increasingly large and complex datasets, and measured how such models perform both on the training dataset (via cross-validation) and on datasets explicitly excluded from the training dataset. Accuracy results are shown in Figure~\ref{fig:model_performances}.  
We note that the addition of datasets about unseen classes of bots in the training data does not seem to degrade model performance: the accuracy remains high across models trained on increasingly richer datasets. In other words, the present framework seems flexible enough to accommodate the learning of new examples without forgetting the old ones.  
Novel datasets used for testing, in general, lead to lower performance compared to when models can learn from those instances. 
But models tend to generalize better when classifying human accounts.  
Bot instances with similar data used for training the model, such as spam bots, also yield better generalization. 
The most significant gain is observed by adding training data about porn bots, which have very few tweets and realistic profile details. 
All models in this experiments are trained using the full set of 1,209 features used in the latest version of Botometer. 
Overall, the current model trained with the new features and on the new datasets improved the cross-validation accuracy from 0.95 to 0.97 AUC. 


\section{Discussion and perspectives}

Research on online manipulation, especially social bot detection, is a rapidly growing field; its contributions are yielding numerous academic publications, tools, and datasets. As long as the arms race between bot creators and bot detection systems continue, this research will remain relevant. 
Beyond the recent research developments, let us make a few considerations on how the online ecosystem may change in the next several years. We focus on the perspectives of social bots and AI countermeasures.

\subsection{Future of social bots} 

In all likelihood, bots will become regular and acknowledged parts of the social media ecosystems of the future. Despite all the nefarious uses of social bots, there are great examples of bots providing valuable services to social media users \citep{monsted2017evidence}. Optimistic viewpoints expect an increase in service bots and regulations that will mitigate the impact of bots with malicious intentions.

Recent developments of deep-learning models create opportunities to employ sequence-to-sequence models for automatic conversations  \citep{li2015diversity,li2016persona}, deep reinforcement learning for controlling account behaviors \citep{mnih2015human,he2016deep,serban2017deep}, and generative adversarial networks for creating fake profile images and content  \citep{goodfellow2014generative,zhao2017dual}. 
Some of these methodologies have already demonstrated a capability to surpass human performance on various tasks, such as face recognition and game playing  \citep{mnih2015human,taigman2014deepface}. 
Deep-learning approaches can also be used to detect sentiment online  \citep{socher2013recursive,tang2014learning,severyn2015twitter}, allowing the next generation of social bots to be more perceptive about human emotions.
Successful implementation and further advances of these technologies can pose serious threats, since discriminating between organic human behavior and automation is likely to become a more and more challenging task, even for experts.


The future of social bots may be shaped by legal, ethical, and political considerations as much as technological ones. Recent events, including election manipulations discussed in Section~\ref{sec:impact}, have led to loud public calls for platforms to apply stricter safeguards against abuse and for policymakers to enact regulations of social media platform in several countries. While platforms are in theory free to police their systems according to their own terms of service and community standards, they are under increasing pressure from politicians. On the one hand, this may lead to needed actions, such as more robust measures to counter impersonation and deception by social bots. On the other hand, political interference could also be counterproductive. For example, aggressive suspension of abusive bots that spread false or misleading political content has been criticized by some politicians as censorship \citep{nytimes2018bias,guardian2018censorship}; even research on bot detection has been targeted by partisan attacks and labeled as suppression of free speech \citep{Mervis686}. 

Another set of legal questions relate to whether bots have rights \citep{cjr2018botfreespeech}.  Are bots entitled to journalistic protections such as those that shield reporters from compelled disclosure of sources? Do they have free-speech rights? The latter question is attracting significant attention following a recent California law requiring online bots to identify themselves as such when they interact with humans, for example promoting a political candidate during elections \citep{politico2018bot1A}. While this law seems to strike a reasonable balance between free speech and regulation, a combination of US Supreme Court decisions about the First Amendment --- the free speech rights of corporations, the right against compelled speech, the right to anonymous speech, and the right to lie --- may lead to the conclusion that states cannot compel entities to disclose the automated nature of online content. These types of questions are yet to be addressed by the courts, and may have huge repercussions on social media platform regulations designed to protect the information ecosystem from manipulation by bots.

\subsection{Future of AI countermeasures} 

So far, the technical response to increasingly sophisticated bots has largely focused on increased sophistication in detection approaches. As reviewed in Section~\ref{sec:review-methods}, recently proposed tools for social bot detection employ deep learning, evolutionary algorithms, and unsupervised learning. These techniques can provide solutions for specific problems, such as addressing the shortage of human-labeled training data on bot activity and analyzing online sentiment. For example, we plan to employ deep-learning models to replace lexicon-based sentiment analysis in the Botometer feature extraction pipeline. 

The detection of coordinated bots will continue to be a challenge. 
Monitoring conversations and coordinated activities among groups of accounts may reveal anomalous behaviors. 
While the unsupervised methods reviewed in Section~\ref{sec:review-methods} have proven effective at identifying some coordinated bots with high precision, they can only check a single feature. As a result, only narrow groups of bots that coordinate in specific ways can be captured. To keep up with all kinds of coordinated manipulation, new algorithms are needed to leverage the many possible dimensions of similarity between suspicious accounts.  
Another research challenge that could be relevant for the dismantling of coordinated bot attacks is the detection of promoted campaigns at their early stages \citep{ferrara2016detection,varol2017early}. Future efforts should aim for a pro-active approach to mitigate the effects of social bots by engaging with them early \citep{cresci2018reaction}. 

Our recent analysis points to the existence of many distinct types of social bots, such as spam bots, mention bots, cyborgs, and many others~\citep{mitter2014categorization}. Botometer employs an ensemble model to learn complex rules able to recognize multiple types of bot behavioral patterns. Detection models need to be able to rapidly adapt to changes in both human and bot behaviors. The iterative process of detecting novel accounts by expanding training datasets, presented in Section~\ref{sec:retraining}, is an effective approach but requires extensive and continual manual efforts. We are currently exploring ways to automatically detect novel bot behavior classes and build specialized classifiers. Using multiple models may lead to better discrimination between human behavior and specific bot strategies.

Machine learning is playing a growing role in decision making on issues of broad societal importance \citep{executive2016big,barocas2016big}. As AI systems employ more sophisticated learning methodologies, model complexity increases and systems become vulnerable to overlooked biases \citep{Kirkpatrick:2016}. Social bot detection systems are not immune to the problem of algorithmic bias. Models trained on manually annotated data might inherit the biases of annotators. 
Algorithmic interpretability would make it easier to spot such biases, and in fact a recent European Union regulation grants users the right to ask for an explanation of an algorithmic decision that was made about them \citep{goodman2016european}. 
Unfortunately, many machine learning algorithms, including those used by Botometer, do not make it easy to interpret their classifications. Deep-learning methods are even more challenging in this respect. As bot detection tools are used to make important decisions, such as whether an account should be suspended, the issue of algorithmic bias is an important direction for future research. 

As discussed in Section~\ref{sec:engagement}, technical solutions should take into account how humans interact with them. Ultimately, the fight against online manipulation must be carried out by our entire society and not just AI researchers. Media literacy efforts may be critical in this respect \citep{lazer2018science}. 

Much public debate revolves around how humans will interact with artificial intelligence. Because of the significant financial and political incentives for developing benign and malicious social bots, this domain provides us with a concrete and realistic scenario to imagine the future of both collaborative and adversarial human-bot interactions. 

\section*{Acknowledgements}

We thank Zoher Kachwala and Shradha Gyaneshwar Baranwal for collecting the celebrity dataset and Gregory Maus for collecting the fake followers dataset. We also thank Andy Patel and Josh Emerson for contributing to the Bot Repository.  
The authors are grateful to their research sponsors, including the Air Force Office of Scientific Research (award  FA9550-17-1-0327) and the Defense Advanced Research Projects Agency (contracts W911NF-12-1-0037 and W911NF-17-C-0094). 
K.-C. Y. was supported in part by the National Institutes of Health (award 5R01DA039928-03). The authors declare no conflicts of interest.

\bibliography{ref}

\end{document}